\documentclass[a4paper,12pt,twoside]{article}
\usepackage{amsmath,amsfonts,amsthm}
\numberwithin{equation}{section} 
\newtheorem{theorem}{Theorem}[section]

\newtheorem{proposition}[theorem]{Proposition}
\newtheorem{lemma}[theorem]{Lemma}
\newtheorem{corollary}[theorem]{Corollary}
\theoremstyle{remark}
\newtheorem{rem}[theorem]{Remark}
\begin{document}
\title{Difference $L$ operators and a 
Casorati determinant solution to the $T$-system 
for twisted quantum affine algebras}
\author{ 
Zengo Tsuboi
\\
{\it Graduate School of Mathematical Sciences, 
University of Tokyo,} \\ 
{\it Komaba 3-8-1, Meguro-ku, Tokyo 153-8914, Japan}
}
\date{}
\maketitle
\begin{abstract}
We propose factorized difference  operators $L(u)$ associated 
with the twisted quantum affine algebras 
$U_{q}(A^{(2)}_{2n}),U_{q}(A^{(2)}_{2n-1}),
U_{q}(D^{(2)}_{n+1}),U_{q}(D^{(3)}_{4})$. 
These operators are shown to be annihilated by a screening operator. 
Based on a basis of the solutions of the difference equation 
$L(u)w(u)=0$, we also construct a Casorati determinant solution to 
 the $T$-system for $U_{q}(A^{(2)}_{2n}),U_{q}(A^{(2)}_{2n-1})$. 
%
\end{abstract}
Journal-ref: J. Phys. A: Math. Gen. 35 (2002) 4363-4373 \\
DOI: 10.1088/0305-4470/35/19/316
\section{Introduction}
In \cite{KS95-2}, a class of functional relations, a $T$-system, 
was proposed for commuting transfer matrices of solvable lattice 
models associated to twisted quantum affine algebras 
$U_{q}(X^{(r)}_{N})$ ($r>1$). 
For $X^{(r)}_{N}=A^{(2)}_{N}$, it has the following form: 

\noindent For $U_{q}(A^{(2)}_{2n})$ case:
\begin{eqnarray}
\hspace{-30pt} && T^{(a)}_{m}(u-1)T^{(a)}_{m}(u+1)=
T^{(a)}_{m-1}(u)T^{(a)}_{m+1}(u)+T^{(a-1)}_{m}(u)T^{(a+1)}_{m}(u) 
\nonumber \\
\hspace{-30pt} && \hspace{180pt} \mathrm{for} \qquad 1 \le a \le n-1,
\label{t-sys1} \\ 
\hspace{-30pt} && T^{(n)}_{m}(u-1)T^{(n)}_{m}(u+1)=
T^{(n)}_{m-1}(u)T^{(n)}_{m+1}(u)+
T^{(n-1)}_{m}(u)T^{(n)}_{m}(u+\frac{\pi i }{2\hbar}).
\nonumber 
\end{eqnarray}
For $U_{q}(A^{(2)}_{2n-1})$ case: 
\begin{eqnarray}
\hspace{-30pt} && T^{(a)}_{m}(u-1)T^{(a)}_{m}(u+1)=
T^{(a)}_{m-1}(u)T^{(a)}_{m+1}(u)+T^{(a-1)}_{m}(u)T^{(a+1)}_{m}(u)
\nonumber \\
\hspace{-30pt} && \hspace{180pt} \mathrm{for} \qquad 1 \le a \le n-1,
\label{t-sys2}
\\ 
\hspace{-30pt} && T^{(n)}_{m}(u-1)T^{(n)}_{m}(u+1)=
T^{(n)}_{m-1}(u)T^{(n)}_{m+1}(u)+
T^{(n-1)}_{m}(u)T^{(n-1)}_{m}(u+\frac{\pi i }{2\hbar}).
\nonumber 
\end{eqnarray}
Here 
$\{T^{(a)}_{m}(u)\}_{a\in I_{\sigma};m\in \mathbb{Z}_{\ge 1};u\in \mathbb{C}}$ 
 ($I_{\sigma}=\{1,2\dots,n\}$)
are the transfer matrices with the auxiliary space labeled by 
$a$ and $m$. We shall adopt the boundary condition 
$T^{(a)}_{-1}(u)=0$, $T^{(a)}_{0}(u)=1$, 
which is natural for the transfer matrices. 
This $T$-system (\ref{t-sys1}), (\ref{t-sys2}) 
is a kind of discrete Toda equation, which follows 
from a reduction 
of the Hirota-Miwa equation \cite{H81},\cite{M82}. 
The original $T$-system \cite{KS95-2} contains 
a scalar function $g^{(a)}_{m}(u)$ in the second term of
 rhs of (\ref{t-sys1}),(\ref{t-sys2}). 
 Throughout this paper, we set $g^{(a)}_{m}(u)=1$. 
This corresponds to the case where the vacuum part is formally trivial. 
However, structure of the solution of 
(\ref{t-sys1}),(\ref{t-sys2}) is essentially independent of 
the function $g^{(a)}_{m}(u)$. 
In this paper, we briefly report on
 a new expression to the solution 
of (\ref{t-sys1}),(\ref{t-sys2}) motivated by the recently found 
interplay \cite{KOSY01} between factorized difference $L$ operators 
and the $q$-characters for non-twisted quantum 
affine algebras \cite{FR99,FM01}. 

In section 2, we propose factorized difference operators $L(u)$ for 
$U_{q}(A^{(2)}_{2n})$, $U_{q}(A^{(2)}_{2n-1})$, 
$U_{q}(D^{(2)}_{n+1})$, $U_{q}(D^{(3)}_{4})$. 
$L(u)$ generates functions 
$\{T^{a}(u)\}_{a \in \mathbb{Z};u \in \mathbb{C}}$, 
which are Laurent 
polynomials in variables $\{Y_{a}(u)\}_{a \in I_{\sigma};
u \in \mathbb{C}}$. 
Moreover $Y_{a}(u)$ is expressed by 
 a function $Q_{a}(u)$ which corresponds to the Baxter $Q$-function. 
When $Q_{a}(u)$ is suitably chosen 
in the context of the analytic Bethe ansatz \cite{R83,R87,KS95,KS95-2}, 
$T^{a}(u)$ corresponds to an eigenvalue formula of the 
transfer matrix in the dressed vacuum form (DVF).  
In particular for $1\le a \le b$ 
($U_{q}(A^{(2)}_{2n}),U_{q}(A^{(2)}_{2n-1})$: $b=n$; 
$U_{q}(D^{(2)}_{n+1})$: $b=n-1$; $U_{q}(D^{(3)}_{4})$: $b=2$), 
 the auxiliary space for this transfer matrix is expected \cite{KS95-2} 
to be a finite dimensional irreducible module of the 
 quantum affine algebra \cite{CP95,CP98}, which is called 
 the Kirillov-Reshetikhin module $W^{(a)}_{1}(u)$ 
 (see also, section 5 in \cite{KNT01}). 
One of the intriguing properties of $L(u)$ is that 
 $L(u)$ is annihilated by a screening operator 
 $\{\mathcal{S}_{a}\}_{a\in I_{\sigma}}$, 
 from which $(\mathcal{S}_{a}\cdot T^{a})(u)=0$ results. 
 In the context of the analytic Bethe ansatz, 
this corresponds to the pole-freeness of $T^{a}(u)$ under the Bethe ansatz 
equation. 
For the non-twisted case $U_{q}(X^{(1)}_{N})$, 
one may identify $\mathcal{S}_{a}$ with 
the Frenkel-Reshetikhin screening operator \cite{FR99} if 
 $Q_{a}(u)$ is suitably chosen. 
 
For $U_{q}(A^{(2)}_{N})$ case, $L(u)$ becomes order of $N+1$. 
By using a basis of the solutions of the difference equation $L(u)w(u)=0$, 
in section 3, we give a solution (Theorem \ref{main}) 
of the $T$-system for 
$U_{q}(A^{(2)}_{N})$ (\ref{t-sys1}), (\ref{t-sys2}) 
as a ratio of two Casorati determinants whose matrix size 
 is constantly $(N+1) \times (N+1)$. On solving this $T$-system,
 a duality relation (Proposition \ref{dual}) plays 
 an important role. 
There is another expression of the solution to the 
$U_{q}(A^{(2)}_{N})$ $T$-system (\ref{t-sys1}), (\ref{t-sys2})
 which is described by semi-standard tableaux with rectangular shape 
\cite{KS95-2}. 
This solution 
 follows from a reduction of the Bazhanov and Reshetikhin's 
Jacobi-Trudi type formula \cite{BR90} (see (\ref{jacobi-trudi})). 
In contrast to the Casorati determinants case, the 
size of the matrix for 
this determinant is $m \times m$ and thus increases 
as $m$ increases. 
Lemma \ref{m-nnsy} connects these two types of solutions. 

In contrast to $U_{q}(A^{(2)}_{N})$ case, $L(u)$ 
for $U_{q}(D^{(2)}_{n+1}),U_{q}(D^{(3)}_{4})$ contain 
factors which have a negative exponent $-1$,  
thus their order become infinite. Therefore we can not 
straightforwardly extend the analysis  to get the Casorati 
determinant type solution for $U_{q}(A^{(2)}_{N})$ to this case. 
However Jacobi-Trudi type formulae are still 
available in this case as reductions of the 
solutions in \cite{KNH96,TK96}. 
This situation is parallel to 
the non-twisted $U_{q}(D^{(1)}_{n})$ case \cite{KOSY01}. 

The deformation parameter $q$ is expressed by 
 a parameter $\hbar$ as $q=\mathrm{e}^{\hbar}$. 
The parameter $\hbar$ often appears as a multiple 
of $\frac{\pi i}{r \hbar}$. 
However we note that our argument in this paper is also valid 
even if one formally set $\frac{\pi i}{r \hbar}=0$. 
In this case, the $T$-system (\ref{t-sys1})
is equivalent to the one for the 
superalgebra $B^{(1)}(0|n)$ \cite{T99}. 

In this paper, we omit most of the calculations and proofs, 
which are parallel with those in the non-twisted case \cite{KOSY01}.  
\section{Difference $L$ Operators}
Let $X_{N}$ be a complex simple Lie algebra 
of rank $N$, $\sigma$ a Dynkin diagram automorphism of $X_{N}$ 
of order $r=1,2,3$. 
The affine Lie algebras of type 
$X^{(r)}_{N}=A^{(1)}_{n}$ ($n\ge 1$), $B^{(1)}_{n}$ ($n\ge 2$), 
$C^{(1)}_{n}$ ($n\ge 2$), 
$D^{(1)}_{n}$ ($n\ge 4$), $E^{(1)}_{n}$ ($n=6,7,8$), 
$F^{(1)}_{4}$, $G^{(1)}_{2}$, 
$A^{(2)}_{2n}$ ($n\ge 1$), $A^{(2)}_{2n-1}$ ($n\ge 2$), 
$D^{(2)}_{n+1}$ ($n\ge 2$), $E^{(2)}_{6}$ 
and $D^{(3)}_{4}$
are realized as the canonical central extension of the 
loop algebras based on the pair $(X_{N},\sigma)$. 
We write the set of the nodes of the Dynkin diagram of 
$X_{N}$ as $I=\{1,2,\dots,N\}$, and let 
$I_{\sigma}=\{1,2,\dots,n\}$ be the set of $\sigma$-orbits of $I$. 
In particular, $N=n$ and $I=I_{\sigma}$ for the non-twisted case $r=1$. 
We define numbers $\{r_{a}\}_{a\in I}$ such that 
$r_{a}=r$ if $\sigma(a)=a$, otherwise $r_{a}=1$. 
%
\begin{figure}[htb]
\caption{
{\small 
The Dynkin diagrams of $X_{N}$ for $r>1$: 
The enumeration of the nodes with $I$ is specified under 
or the right side of the nodes. 
The filled circles denote the fixed points of the Dynkin diagram 
automorphism $\sigma$ of order $r$.
}}
\unitlength=1.4pt
\label{dynkin}
\begin{tabular}[t]{l|c|l}
$X^{(r)}_{N}$ & $X_{N}$ & automorphism $\sigma$ \\
\hline 
$A^{(2)}_{2n}$ &
\begin{picture}(112,20)(-5,-5)
\multiput( 0,0)(20,0){2}{\circle{6}}
\multiput(40,0)(20,0){2}{\circle{6}}
\multiput(80,0)(20,0){2}{\circle{6}}
\multiput( 3,0)(20,0){1}{\line(1,0){14}}
\multiput(23,0)(4,0){4}{\line(1,0){2}}
\multiput(43,0)(20,0){1}{\line(1,0){14}}
\multiput(63,0)(4,0){4}{\line(1,0){2}}
\multiput(83,0)(20,0){1}{\line(1,0){14}}
\put(0,-6){\makebox(0,0)[t]{\scriptsize{$1$}}}
\put(20,-6){\makebox(0,0)[t]{\scriptsize{$2$}}}
\put(40,-7){\makebox(0,0)[t]{\tiny{$n$}}}
\put(60,-6){\makebox(0,0)[t]{\tiny{$n \! + \! 1$}}}
\put(80,-6){\makebox(0,0)[t]{\tiny{$2n \! - \! 1$}}}
\put(100,-6){\makebox(0,0)[t]{\tiny{$2n $}}}
\end{picture}
& {\footnotesize$\sigma(2n-a+1)=a$ for $1\le a \le 2n$}
\\
$A^{(2)}_{2n-1}$&
\begin{picture}(112,20)(-5,-5)
\multiput( 0,0)(20,0){2}{\circle{6}}
\multiput(50,0)(20,0){1}{\circle*{6}}
\multiput(80,0)(20,0){2}{\circle{6}}
\multiput( 3,0)(20,0){1}{\line(1,0){14}}
\multiput(83,0)(20,0){1}{\line(1,0){14}}
\multiput(23,0)(4,0){6}{\line(1,0){2}}
\multiput(53,0)(4,0){6}{\line(1,0){2}}
\put(0,-6){\makebox(0,0)[t]{\scriptsize{$1$}}}
\put(20,-6){\makebox(0,0)[t]{\scriptsize{$2$}}}
\put(50,-7){\makebox(0,0)[t]{\tiny{$n$}}}
\put(80,-6){\makebox(0,0)[t]{\tiny{$2n \! -\! 2$}}}
\put(100,-6){\makebox(0,0)[t]{\tiny{$2n \! - \! 1 $}}}
\end{picture}
& {\footnotesize$\sigma(2n-a)=a$ for $1\le a \le 2n-1$}
\\
$D^{(2)}_{n+1}$&
\begin{picture}(112,40)(-5,-5)
\multiput( 0,0)(20,0){2}{\circle*{6}}
\multiput(80,0)(20,0){1}{\circle*{6}}
\multiput(100,0)(20,0){1}{\circle{6}}
\put(80,20){\circle{6}}
\multiput( 3,0)(20,0){2}{\line(1,0){14}}
\multiput(63,0)(20,0){2}{\line(1,0){14}}
\multiput(39,0)(4,0){6}{\line(1,0){2}}
\put(80,3){\line(0,1){14}}
\put(0,-6){\makebox(0,0)[t]{\scriptsize{$1$}}}
\put(20,-6){\makebox(0,0)[t]{\scriptsize{$2$}}}
\put(80,-6){\makebox(0,0)[t]{\tiny{$n \! - \! 1$}}}
\put(103,-6){\makebox(0,0)[t]{\tiny{$n \! + \! 1$}}}
\put(85,20){\makebox(0,0)[l]{\tiny{$n$}}}
\end{picture}
&
{\footnotesize 
$\begin{array}{l}
\sigma(a)=a \;\; {\rm for} \;\; 1\le a \le n-1;\\ 
\sigma(n)=n+1; \;\; \sigma(n+1)=n
\end{array}$ 
}
\\
$E^{(2)}_{6}$&
\begin{picture}(90,40)(-5,-5)
\multiput(0,0)(20,0){2}{\circle{6}}
\multiput(40,0)(20,0){1}{\circle*{6}}
\multiput(60,0)(20,0){2}{\circle{6}}
\put(40,20){\circle*{6}}
\multiput(3,0)(20,0){4}{\line(1,0){14}}
\put(40, 3){\line(0,1){14}}
\put( 0,-6){\makebox(0,0)[t]{\scriptsize{$1$}}}
\put(20,-6){\makebox(0,0)[t]{\scriptsize{$2$}}}
\put(40,-6){\makebox(0,0)[t]{\scriptsize{$3$}}}
\put(60,-6){\makebox(0,0)[t]{\scriptsize{$5$}}}
\put(80,-6){\makebox(0,0)[t]{\scriptsize{$6$}}}
\put(46,20){\makebox(0,0)[l]{\scriptsize{$4$}}}
\end{picture}
&
{\footnotesize 
$\begin{array}{l}
\sigma(7-a)=a \;\; {\rm for} \;\; a=1,2,5,6; \\
\sigma(3)=3; \;\; \sigma(4)=4
\end{array}$ 
}

\\
$D^{(3)}_{4}$&
\begin{picture}(50,40)(-5,-5)
\multiput(0,0)(20,0){1}{\circle{6}}
\multiput(20,0)(20,0){1}{\circle*{6}}
\multiput(40,0)(20,0){1}{\circle{6}}
\put(20,20){\circle{6}}
\multiput(3,0)(20,0){2}{\line(1,0){14}}
\put(20, 3){\line(0,1){14}}
\put( 0,-6){\makebox(0,0)[t]{\scriptsize{$1$}}}
\put(20,-6){\makebox(0,0)[t]{\scriptsize{$2$}}}
\put(40,-6){\makebox(0,0)[t]{\scriptsize{$4$}}}
\put(26,20){\makebox(0,0)[l]{\scriptsize{$3$}}}
\end{picture}
& {\footnotesize
$\begin{array}{l}
\sigma(1)=3; \;\; \sigma(2)=2; \\ 
\sigma(3)=4; \;\; \sigma(4)=1
\end{array}
$}
\\
\end{tabular}
\end{figure}
 In our enumeration 
of the notes of the Dynkin diagram (see, Figure \ref{dynkin}), 
$r_{a}$ is $1$ except for the case: 
$r_{n}=2$ for $A^{(2)}_{2n-1}$, 
$r_{a}=2$ ($1\le a \le n-1$) for $D^{(2)}_{n+1}$, 
$r_{3}=r_{4}=2$ for $E^{(2)}_{6}$, 
$r_{2}=3$ for $D^{(3)}_{4}$. 
Let $\{\alpha_{a}\}_{a\in I}$ be the simple roots of $X_{N}$ with 
a bilinear form $(\cdot |\cdot)$ normalized as $(\alpha|\alpha)=2$ 
for a long root $\alpha$. 
Let $I_{ab}$ be an element of the incidence matrix of $X_{N}$: 
$I_{ab}=2\delta_{ab}-2(\alpha_{a}|\alpha_{b})/(\alpha_{a}|\alpha_{a})$. 

Let $U_{q}(X^{(r)}_{N})$ be the quantum affine algebra. 
 We introduce functions 
 $\{Q_{a}(u)\}_{a \in I_{\sigma};u\in \mathbb{C}}$ which correspond to 
 the Baxter $Q$ functions for $U_{q}(X^{(r)}_{N})$, 
 and define functions 
 $\{Y_{a}(u)\}_{a \in I_{\sigma};u\in \mathbb{C}}$ as  
\begin{eqnarray}
Y_{a}(u)=\frac{Q_{a}(u-\frac{(\alpha_{a}|\alpha_{a})}{2})}
{Q_{a}(u+\frac{(\alpha_{a}|\alpha_{a})}{2})}.
\end{eqnarray}
We formally set 
$Y_{0}(u)=1$;  
$Q_{n+1}(u)=Q_{n}(u+\frac{\pi i}{2 \hbar})$ and 
$Y_{n+1}(u)=Y_{n}(u+\frac{\pi i}{2 \hbar})$ 
for 
$X^{(r)}_{N}=A^{(2)}_{2n}$; 
$Q_{n+1}(u)=1$ and $Y_{n+1}(u)=1$ for $X^{(r)}_{N} \ne A^{(2)}_{2n}$. 
For the twisted case $r>1$, 
we assume quasi-periodicity 
$Q_{a}(u+\frac{\pi i}{\hbar})=h_{a}Q_{a}(u)$ ($h_{a}\in \mathbb{C}$),
 which induces periodicity
 $Y_{a}(u+\frac{\pi i}{\hbar})=Y_{a}(u)$. 
For the non-twisted case $r=1$, 
one can identify $Y_{a}(u)$ with the 
Frenkel-Reshetikhin variable $Y_{a,q^{u}}$ \cite{FR99} 
 denoted as $Y_{a}(u)$ in \cite{KOSY01} if 
 $Q_{a}(u)$ is suitably chosen. 
We shall also use notations 
$Q^{k}_{a}(u)=\prod_{j=0}^{k-1}Q_{a}(u+\frac{\pi ij}{r\hbar})$ and 
$Y^{k}_{a}(u)=\prod_{j=0}^{k-1}Y_{a}(u+\frac{\pi ij}{r\hbar})$. 

Next we introduce screening operators 
$\{\mathcal{S}_{a}\}_{a\in I_{\sigma}}$ 
on $\mathbb{Z}[Y_{a}(u)^{\pm 1}]_{a\in I_{\sigma};u\in \mathbb{C}}$, 
whose action is given by 
\begin{eqnarray}
({\mathcal S}_{a} \cdot Y_{b})(u)=\delta_{ab}Y_{a}(u)S_{a}(u).
\end{eqnarray}
Here we assume $S_{a}(u)$ satisfies the following relation 
\begin{eqnarray}
&& S_{a}\left(u+(\alpha_{a}|\alpha_{a})\right)=
A_{a}\left(u+\frac{(\alpha_{a}|\alpha_{a})}{2}\right)S_{a}(u),
\label{sc-eq} \\
&& 
A_{a}(u)=\prod_{b=1}^{n^{\prime}}
\frac{Q_{b}^{r_{ab}}(u-(\alpha_{a}|\alpha_{b}))}
{Q_{b}^{r_{ab}}(u+(\alpha_{a}|\alpha_{b}))},
\label{mbae}
\end{eqnarray}
where $r_{ab}=\max(r_{a},r_{b})$; 
$n^{\prime}=n+1$ for $X^{(r)}_{N}=A^{(2)}_{2n}$ 
and $n^{\prime}=n$ for $X^{(r)}_{N} \ne A^{(2)}_{2n}$. 
We assume $\mathcal{S}_{a}$ obeys the Leibniz rule. 
The origin of (\ref{mbae}) goes back to the 
Reshetikhin and Wiegmann's 
Bethe ansatz equation \cite{RW87} (cf. (\ref{BAE})). 
For the non-twisted case $r=1$ case, (\ref{mbae}) 
reduces to the corresponding equation in \cite{KOSY01}. 
We have a formal 
solution of (\ref{sc-eq}) (see also, section 5 in \cite{FR99}): 
\begin{eqnarray}
S_{a}(u)=\frac{\prod_{b=1}^{n^{\prime}}K_{ab}(u)}
{Q^{r_{a}}_{a}(u-\frac{(\alpha_{a}|\alpha_{a})}{2})
Q^{r_{a}}_{a}(u+\frac{(\alpha_{a}|\alpha_{a})}{2})},
\end{eqnarray}
where 
\begin{eqnarray}
K_{ab}(u)=
\left\{
 \begin{array}{lll} 
 1  & \mathrm{if} & I_{ab}=0 \\
 Q^{r_{ab}}_{b}(u) & \mathrm{if} & I_{ab}=1 \\
 Q_{b}(u-\frac{1}{2})Q_{b}(u+\frac{1}{2}) & 
 \mathrm{if} & I_{ab}=2 \\
 Q_{b}(u-\frac{2}{3})Q_{b}(u)Q_{b}(u+\frac{2}{3}) 
 & \mathrm{if} & I_{ab}=3 .
 \end{array}
\right.
\end{eqnarray}
Owing to the Leibniz rule, we have
\begin{eqnarray}
({\mathcal S}_{a} \cdot Y^{k}_{b})(u)=
\delta_{ab}Y^{k}_{a}(u) \sum_{j=0}^{k-1}S_{a}(u+\frac{\pi ij}{r\hbar}).
\end{eqnarray}
We shall use the following variables for each algebra; 
the origin of these variables goes back to the analytic 
Bethe ansatz calculation of DVF \cite{R87,KS95,KS95-2}. \\
For $U_{q}(A^{(2)}_{2n})$ case:
\begin{eqnarray}
&& z_{a}(u)=\frac{Y_{a}(u+a)}{Y_{a-1}(u+a+1)} 
\quad \mathrm{for} \quad 1 \le a \le n, \nonumber \\
&& z_{0}(u)=
\frac{Y_{n}(u+n+1+\frac{\pi i }{2\hbar})}{Y_{n}(u+n+2)},
\label{z-a22n} \\ 
&& z_{\overline{a}}(u)=
\frac{Y_{a-1}(u+2n-a+2+\frac{\pi i }{2\hbar})}
     {Y_{a}(u+2n-a+3+\frac{\pi i }{2\hbar})} 
\quad \mathrm{for} \quad
 1 \le a \le n. \nonumber 
\end{eqnarray}
We also use the variables: 
$x_{a}(u)=z_{a}(u)$ and 
$x_{2n-a+2}(u)=z_{\overline{a}}(u)$ for $1 \le a \le n$; 
$x_{n+1}(u)=z_{0}(u)$.\\
For $U_{q}(A^{(2)}_{2n-1})$ case: 
\begin{eqnarray}
&& z_{a}(u)=\frac{Y_{a}(u+a)}{Y_{a-1}(u+a+1)} 
\quad \mathrm{for} \quad 
1 \le a \le n-1, \nonumber \\ 
&& z_{n}(u)=
\frac{Y_{n}^{2}(u+n)}{Y_{n-1}(u+n+1)}, 
\nonumber \\ 
&& z_{\overline{n}}(u)=
\frac{Y_{n-1}(u+n+1+\frac{\pi i }{2\hbar})}
{Y_{n}^{2}(u+n+2)}, 
\label{z-a22n-1} \\
&& z_{\overline{a}}(u)=
\frac{Y_{a-1}(u+2n-a+1+\frac{\pi i }{2\hbar})}
     {Y_{a}(u+2n-a+2+\frac{\pi i }{2\hbar})} 
\quad \mathrm{for} \quad 1 \le a \le n-1. \nonumber 
\end{eqnarray}
We also use the variables: 
$x_{a}(u)=z_{a}(u)$ and 
$x_{2n-a+1}(u)=z_{\overline{a}}(u)$ for $1 \le a \le n$.\\
For $U_{q}(D^{(2)}_{n+1})$ case:
\begin{eqnarray}
&& z_{a}(u)=\frac{Y_{a}^{2}(u+a)}{Y_{a-1}^{2}(u+a+1)} 
\quad \mathrm{for} \quad 
1 \le a \le n, \nonumber \\
&& z_{n+1}(u)=
\frac{Y_{n}(u+n+\frac{\pi i }{2\hbar})}{Y_{n}(u+n+2)}, 
\nonumber \\ 
&& z_{\overline{n+1}}(u)=
\frac{Y_{n}(u+n)}{Y_{n}(u+n+2+\frac{\pi i }{2\hbar})}, 
\label{z-d2n+1} \\ 
&& z_{\overline{a}}(u)=
\frac{Y_{a-1}^{2}(u+2n-a+1)}
     {Y_{a}^{2}(u+2n-a+2)} 
\quad \mathrm{for} \quad
 1 \le a \le n. \nonumber 
\end{eqnarray}
For $U_{q}(D^{(3)}_{4})$ case:
\begin{eqnarray}
&& z_{1}(u)=Y_{1}(u+1), \nonumber \\
&& z_{2}(u)=\frac{Y_{2}^{3}(u+2)}{Y_{1}(u+3)}, \nonumber \\
&& z_{3}(u)=\frac{Y_{1}^{3}(u+3)}{Y_{1}(u+3)Y_{2}^{3}(u+4)},
 \nonumber \\
&& z_{4}(u)=\frac{Y_{1}(u+3-\frac{\pi i }{3\hbar})}
                 {Y_{1}(u+5+\frac{\pi i }{3\hbar})}, \nonumber \\
&& z_{\overline{4}}(u)=\frac{Y_{1}(u+3+\frac{\pi i }{3\hbar})}
                 {Y_{1}(u+5-\frac{\pi i }{3\hbar})}, \label{z-d34} \\
&& z_{\overline{3}}(u)=\frac{Y_{1}(u+5)Y_{2}^{3}(u+4)}{Y_{1}^{3}(u+5)},
\nonumber \\
&& z_{\overline{2}}(u)=\frac{Y_{1}(u+5)}{Y_{2}^{3}(u+6)},\nonumber \\
&& z_{\overline{1}}(u)=\frac{1}{Y_{1}(u+7)}.\nonumber 
\end{eqnarray}
Let $D$ be a difference operator such that 
$Df(u)=f(u+2)D$ for any function $f(u)$. 
We shall use notations: 
$\overrightarrow{\prod_{k=1}^{m}}g_{k}=g_{1}g_{2}\cdots g_{m}$
and 
$\overleftarrow{\prod_{k=1}^{m}}g_{k}=g_{m}g_{m-1}\cdots g_{1}$. 
By using the variables (\ref{z-a22n})-(\ref{z-d34}), 
we introduce a factorized difference $L$ operator for 
each algebra. \\
For $U_{q}(A^{(2)}_{2n})$ case:
\begin{eqnarray}
L(u)&=&
\overrightarrow{\prod_{a=1}^{n}}(1-z_{\overline{a}}(u)D)
(1-z_{0}(u)D)
\overleftarrow{\prod_{a=1}^{n}}(1-z_{a}(u)D) \nonumber \\ 
&=&
\overleftarrow{\prod_{a=1}^{2n+1}}(1-x_{a}(u)D).
\label{L-a22n}
\end{eqnarray}
For $U_{q}(A^{(2)}_{2n-1})$ case:
\begin{eqnarray}
L(u)=
\overrightarrow{\prod_{a=1}^{n}}(1-z_{\overline{a}}(u)D)
\overleftarrow{\prod_{a=1}^{n}}(1-z_{a}(u)D)=
\overleftarrow{\prod_{a=1}^{2n}}(1-x_{a}(u)D).
\label{L-a22n-1}
\end{eqnarray}
For $U_{q}(D^{(2)}_{n+1})$ case:
\begin{eqnarray}
\hspace{-10pt} 
L(u)=
\overrightarrow{\prod_{a=1}^{n+1}}(1-z_{\overline{a}}(u)D)
(1-z_{n+1}(u)z_{\overline{n+1}}(u+2)D^{2})^{-1}
\overleftarrow{\prod_{a=1}^{n+1}}(1-z_{a}(u)D).
\label{L-d2n+1}
\end{eqnarray}
For $U_{q}(D^{(3)}_{4})$ case:
\begin{eqnarray}
L(u)=
\overrightarrow{\prod_{a=1}^{4}}(1-z_{\overline{a}}(u)D)
(1-z_{4}(u)z_{\overline{4}}(u+2)D^{2})^{-1}
\overleftarrow{\prod_{a=1}^{4}}(1-z_{a}(u)D).
\label{L-d34}
\end{eqnarray}
In general, $L(u)$ (\ref{L-a22n})-(\ref{L-d34})
 are power series of $D$ whose 
coefficients lie in 
$\mathbb{Z}[Y_{a}(u)^{\pm 1}]_{a\in I_{\sigma};u\in \mathbb{C}}$. 
We assume $\mathcal{S}_{a}$ acts on these coefficients linearly. 
\begin{proposition}\label{SL}
For $a \in I_{\sigma}$, we have $({\mathcal S}_{a} \cdot L)(u)=0 $.
\end{proposition}
The proof is similar to the non-twisted case \cite{KOSY01}. 
So we just 
mention the lemmas which are necessary to $U_{q}(D^{(3)}_{4})$ case. 
\begin{lemma}
For $U_{q}(D^{(3)}_{4})$ case, let 
\begin{eqnarray*}
 && H_{1}(u)=Y_{1}(u)+\frac{Y_{2}^{3}(u+1)}{Y_{1}(u+2)}, \quad 
 H_{2}(u)=Y_{2}^{3}(u)+\frac{Y_{1}^{3}(u+1)}{Y_{2}^{3}(u+2)}, \\
&& K_{1}(u)=\frac{1}{Y_{1}(u)}+\frac{Y_{1}(u-2)}{Y_{2}^{3}(u-1)}, \quad 
 K_{2}(u)=\frac{1}{Y_{2}^{3}(u)}+\frac{Y_{2}^{3}(u-2)}{Y_{1}^{3}(u-1)},
\end{eqnarray*}
then $(\mathcal{S}_{a}\cdot H_{a})(u)=(\mathcal{S}_{a}\cdot K_{a})(u)=0$ 
for $a=1,2$. 
\end{lemma}
\begin{lemma}
For $U_{q}(D^{(3)}_{4})$ case, one can rewrite $L(u)$ 
(\ref{L-d34}) as follows:
\begin{eqnarray*}
\hspace{-28pt} && L(u)=
(1-K_{1}(u+7)D+\frac{1}{Y_{2}^{3}(u+8)}D^{2})  \\ 
\hspace{-28pt} &&\times 
(1-\sum_{j=0}^{\infty}A_{j}(u)D^{2j+1}+\sum_{j=0}^{\infty}B_{j}(u)D^{2j+2}) 
(1-H_{1}(u+1)D+Y_{2}^{3}(u+2)D^{2}), \nonumber 
\end{eqnarray*}
where 
\begin{eqnarray*}
\hspace{-28pt} && A_{j}(u)=K_{1}(u+4j+5+\frac{\pi i}{3\hbar})
 H_{1}(u+3-\frac{\pi i}{3\hbar}) \nonumber \\
\hspace{-28pt} && \hspace{50pt} + 
 (1-\delta_{j0})K_{1}(u+4j+5-\frac{\pi i}{3\hbar})
 H_{1}(u+3+\frac{\pi i}{3\hbar}), \nonumber \\
\hspace{-28pt} && B_{j}(u)=K_{1}(u+4j+7+\frac{\pi i}{3\hbar})
 H_{1}(u+3+\frac{\pi i}{3\hbar}) \nonumber \\
\hspace{-28pt} && \hspace{50pt} + 
 K_{1}(u+4j+7-\frac{\pi i}{3\hbar})
 H_{1}(u+3-\frac{\pi i}{3\hbar})
 -\delta_{j0} \frac{Y_{2}^{3}(u+4)}{Y_{2}^{3}(u+6)}.
\end{eqnarray*}
\end{lemma}
\begin{lemma}
For $U_{q}(D^{(3)}_{4})$ case, one can expand the $Y_{2}$ dependent part in 
$L(u)$ (\ref{L-d34}):
\begin{eqnarray*}
&&(1-z_{\overline{2}}(u)D)(1-z_{\overline{3}}(u)D)
=1-Y_{1}(u+5)K_{2}(u+6)D \nonumber \\ 
&& \hspace{160pt} +\frac{Y_{1}(u+5)Y_{1}(u+7)}{Y_{1}^{3}(u+7)}D^{2},
\\
&&(1-z_{3}(u)D)(1-z_{2}(u)D)
=1-\frac{H_{2}(u+2)}{Y_{1}(u+3)}D+
\frac{Y_{1}^{3}(u+3)}{Y_{1}(u+3)Y_{1}(u+5)}D^{2}.
\nonumber 
\end{eqnarray*}
\end{lemma}
We shall expand $L(u)$ as 
\begin{eqnarray}
L(u)=\sum_{a=0}^{\infty}(-1)^{a}T^{a}(u+a)D^{a}. 
\label{expand-L}
\end{eqnarray}
In particular, 
we have $T^{0}(u)=1$ and $T^{a}(u)=0$ for $a \in {\mathbb Z}_{<0}$. 
For $U_{q}(A^{(2)}_{N})$ case, (\ref{expand-L}) becomes a polynomial in $D$ 
of order $N+1$ 
and $T^{a}(u)=0$ for $a \in {\mathbb Z}_{\ge N+2}$. 
\begin{rem}
There is a homomorphism $\beta$ analogous to the one in \cite{FR99}.
\begin{eqnarray*}
\beta :
 {\mathbb Z}[Y_{a}(u)^{\pm 1}]_{a\in I_{\sigma};u\in {\mathbb C}}
 \to 
 {\mathbb Z}[{\mathrm e}^{\pm \frac{1}{r_{a}}
  \Lambda_{a}}]_{a\in I_{\sigma}}; \quad 
  \beta(Y_{a}(u)^{\pm 1})={\mathrm e}^{\pm \frac{1}{r_{a}}\Lambda_{a}},
\end{eqnarray*} 
where $\{\Lambda_{a}\}_{a \in I_{\sigma}}$ are the fundamental weights 
of a rank $n$ subalgebra $\overset{\circ}{\mathfrak{g}}$ 
of $X^{(r)}_{N}$: 
($X^{(r)}_{N},\overset{\circ}{\mathfrak{g}}$)=
($X^{(1)}_{n},X_{n}$), ($A^{(2)}_{2n},C_{n}$), 
($A^{(2)}_{2n-1},C_{n}$), ($D^{(2)}_{n+1},B_{n}$), 
($D^{(3)}_{4},G_{2}$), ($E^{(2)}_{6},F_{4}$). 
Note that the image of $\beta$ is 
independent of the parameter $\hbar$. 
In particular, $\beta(T^{a}(u)) \in 
{\mathbb Z}[{\mathrm e}^{\pm \Lambda_{b}}]_{b\in I_{\sigma}}$ 
is a linear combination of 
$\overset{\circ}{\mathfrak{g}}$ characters 
(cf. section 6 in \cite{HKOTT01}). 
For $1\le a \le b$ 
($U_{q}(A^{(2)}_{2n}),U_{q}(A^{(2)}_{2n-1})$: $b=n$; 
$U_{q}(D^{(2)}_{n+1})$: $b=n-1$; $U_{q}(D^{(3)}_{4})$: $b=2$),
$T^{a}(u)$ contains a term 
$Y_{a}^{r_{a}}(u)=\prod_{k=1}^{a}z_{k}(u+a-2k)$: 
$\beta(Y_{a}^{r_{a}}(u))={\mathrm e}^{\Lambda_{a}}$. 
In the context  of the analytic Bethe ansatz \cite{KS95}
(resp. the theory of $q$-characters \cite{FR99}), 
$Y_{a}^{r_{a}}(u)$ corresponds to  
the top term of DVF 
(resp. the highest weight monomial of the $q$-character)
for the Kirillov-Reshetikhin module $W^{(a)}_{1}(u)$ 
over $U_{q}(X_{N}^{(r)})$. 
\end{rem}
From the Proposition \ref{SL}, we obtain: 
\begin{corollary}
For $a \in I_{\sigma}$ and $b\in {\mathbb Z}$, 
we have $({\mathcal S}_{a} \cdot T^{b})(u)=0 $.
\end{corollary}
For $U_{q}(A^{(2)}_{N})$ case, 
there is a duality among $\{T^{a}(u)\}_{a\in \mathbb{Z};u\in \mathbb{C}}$. 
\begin{proposition}\label{dual}
For $U_{q}(A^{(2)}_{N})$ case, we have 
\begin{eqnarray*}
T^{a}(u)=T^{N+1-a}(u+\frac{\pi i}{2\hbar}), 
\quad a \in {\mathbb Z}.
\end{eqnarray*}
\end{proposition}
This relation is given in \cite{KS95-2}
as \symbol{"60}modulo $\sigma$ relation\symbol{"27}. 
The proof of this proposition is similar 
to the $B^{(1)}(0|n)$ case \cite{T99},
which corresponds to $N=2n$ and $\frac{\pi i}{\hbar} \to 0$. 

One can show
\begin{eqnarray}
L(u)Q_{1}^{r_{1}}(u)=0 \label{LQ}.
\end{eqnarray}
A $T-Q$ relation follows from (\ref{LQ}): 
\begin{eqnarray}
\sum_{a=0}^{\infty}(-1)^{a}T^{a}(u+a)Q_{1}^{r_{1}}(u+2a)=0. 
\label{TQ}
\end{eqnarray}
We shall expand $L(u)^{-1}$ as 
\begin{eqnarray}
L(u)^{-1}=\sum_{m=0}^{\infty}T_{m}(u+m)D^{m}. 
\label{expand-L-yoko}
\end{eqnarray}
In particular, 
we have $T_{0}(u)=1$ and $T_{m}(u)=0$ for $m \in {\mathbb Z}_{<0}$. 
From the relation $L(u)L(u)^{-1}=1$, 
we obtain a $T-T$ relation 
\begin{eqnarray}
\sum_{a=0}^{m}(-1)^{a}T_{m-a}(u+m+a)T^{a}(u+a)=\delta_{m0} .
\label{TT-1}
\end{eqnarray}
From the relation $L(u)^{-1}L(u)=1$, we also have
\begin{eqnarray}
\sum_{a=0}^{m}(-1)^{a}T_{m-a}(u-m-a)T^{a}(u-a)=\delta_{m0}. 
\label{TT-2}
\end{eqnarray}
In particular for $U_{q}(A^{(2)}_{N})$ case, the $T-Q$ relation (\ref{TQ}) 
reduces to 
\begin{eqnarray}
\sum_{a=0}^{N+1}(-1)^{a}T^{a}(u+a)Q_{1}(u+2a)=0.
 \label{TQ-1}
\end{eqnarray}
From the Proposition \ref{dual}, one can rewrite this as follows
\begin{eqnarray}
\sum_{a=0}^{N+1}(-1)^{a}
 T^{a}(u-a)Q_{1}(u-2a+g+\frac{\pi i}{2\hbar})=0,
  \label{TQ-2}
\end{eqnarray}
where $g=N+1$ is the dual Coxeter number of $A^{(2)}_{N}$. 
If one assume $\lim_{m \to \infty}T_{m}(u+m)$ 
(resp. $\lim_{m \to \infty}T_{m}(u-m)$)
is proportional to $Q_{1}(u)$ (resp. $Q_{1}(u+g+\frac{\pi i}{2\hbar})$), 
then one can recover the $T-Q$ relation (\ref{TQ-1})
 (resp. (\ref{TQ-2})) from 
the $T-T$ relation (\ref{TT-1}) (resp. (\ref{TT-2})). 
\section{Solution of the $T$-system}
The goal of this section is to 
give a Casorati determinant solution to 
the $U_{q}(A^{(2)}_{N})$ $T$-system (\ref{t-sys1}),(\ref{t-sys2}). 
Consider the following difference equation 
\begin{eqnarray}
L(u)w(u)=0,\label{d-eq}
\end{eqnarray}
where $L(u)$ is the difference $L$ operator 
(\ref{L-a22n}) and (\ref{L-a22n-1})
for $U_{q}(A^{(2)}_{N})$. By using a basis 
 $\{w_{1}(u),w_{2}(u),\dots, w_{N+1}(u)\}$ of the solutions 
of (\ref{d-eq}), we define a Casorati determinant:
\begin{eqnarray}
[i_{1},i_{2},\dots,i_{N+1}]=
\begin{array}{|cccc|}
w_{1}(u+2i_{1}) & w_{1}(u+2i_{2}) & \cdots & w_{1}(u+2i_{N+1}) \\
w_{2}(u+2i_{1}) & w_{2}(u+2i_{2}) & \cdots & w_{2}(u+2i_{N+1}) \\
\vdots & \vdots & \ddots & \vdots \\
w_{N+1}(u+2i_{1}) & w_{N+1}(u+2i_{2}) & \cdots & w_{N+1}(u+2i_{N+1}) 
\end{array}.
\nonumber 
\end{eqnarray}
Setting $w=w_{1},w_{2},\dots,w_{N+1}$ in (\ref{d-eq}) and
noting the relation $T^{N+1}(u)=1$, 
we obtain the following relation:
\begin{eqnarray}
 [0,1,\dots,N]=[1,2,\dots,N+1].\label{shift}
\end{eqnarray}
Owing to the Cramer's formula, we also have:
\begin{proposition}\label{ta1}
For $a\in \{0,1,\dots,N+1 \}$, we have 
\begin{eqnarray*}
 T^{a}(u+a)=\frac{[0,1,\dots,a-1,a+1,\dots,N+1]}{[0,1,\dots,N]}.
\end{eqnarray*}
\end{proposition}
\begin{lemma}
For $U_{q}(A^{(2)}_{N})$ case, one can rewrite $L(u)$ 
(\ref{L-a22n}),(\ref{L-a22n-1}) as 
\begin{eqnarray*}
L(u)=\overrightarrow{
\prod_{a=1}^{N+1}}(x_{a}(u+N+1-2a+\frac{\pi i}{2\hbar})-D).
\end{eqnarray*}
\end{lemma}
Let $\xi^{(a)}_{m}(u)=[0,1,\dots,a-1,a+m,a+m+1,\dots,N+m]$ and 
$\xi(u)=\xi^{(1)}_{0}(u)=[0,1,\dots,N]$. 
Note that $\xi^{(0)}_{m}(u)=\xi(u)$ follows from (\ref{shift}). 
For $1\le a \le N+1$, we introduce a difference operator
\begin{eqnarray}
L_{a}(u)=\overrightarrow{
\prod_{b=N+2-a}^{N+1}}(D-x_{b}(u+N+1-2b+\frac{\pi i}{2\hbar})).
\end{eqnarray}
In particular we have $L_{N+1}(u)=(-1)^{N+1}L(u)$. 
We choose a basis of the solutions of (\ref{d-eq}) so that it satisfies 
$L_{a}(u)w_{b}(u)=0$ for $1\le b \le a \le N+1$:  
$w_{a} \in \mathrm{Ker}L_{a}$.
For this basis, the following lemma hold. 
\begin{lemma}\label{m-nnsy}
Let $\{i_{k}\}$ be integers such that 
$0=i_{0}<i_{1}<\cdots <i_{N}$, $\mu=(\mu_{k})$ the Young diagram
whose $k$-th row is $\mu_{k}=i_{N+1-k}+k-N-1$, 
and $\mu^{\prime}=(\mu^{\prime}_{k})$ the transposition of $\mu$.
 We assign coordinates $(j,k)\in \mathbb{ Z}^{2}$ 
on the skew-Young diagram $(\mu_{1}^{N+1})/\mu $
such that the row index $j$ increases as we go upwards and the column 
index $k$ increases as we go from the left to the right and that 
$(1,1)$ is on the bottom left corner of $(\mu_{1}^{N+1})/\mu$.
\begin{eqnarray*}
 \frac{[i_{0},i_{1},\dots,i_{N}]}{[0,1,\dots,N]}&=&
 \sum_{b} \prod_{(j,k)\in (\mu_{1}^{N+1})/\mu }
 x_{b(j,k)}(u+2j+2k-4) \\
 &=&\det_{1 \le j,k \le \mu_{1}}
 (T^{\mu^{\prime}_{j}-j+k}
 (u+N-1+j+k-\mu^{\prime}_{j}+\frac{\pi i }{2\hbar})),
\end{eqnarray*}
where the summation is taken over the semi-standard tableau $b$ 
on the skew-Young diagram $(\mu_{1}^{N+1})/\mu$ as the set of elements 
$b(j,k)\in \{1,2,\dots,N+1 \}$ 
 labeled by the coordinates $(j,k)$ mentioned above. 
\end{lemma}
The proof is similar to the $U_{q}(C^{(1)}_{n})$ case \cite{KOSY01}, 
where we use 
a theorem in \cite{NNSY00} and Proposition \ref{dual}. 
Note that Lemma \ref{m-nnsy} reduces to the Proposition \ref{ta1} 
if we set $i_{b}=b$ for $0\le b \le a-1$ and $i_{b}=b+1$ for 
$a \le b \le N$. 
From Proposition \ref{dual} and Lemma \ref{m-nnsy}, one can show: 
\begin{lemma}\label{dual-Ca}
For $a \in \{0,1,\dots,N+1\}$, we have 
$\frac{\xi^{(a)}_{m}(u)}{\xi(u)}=
\frac{\xi^{(N-a+1)}_{m}(u+2a-N-1+\frac{\pi i }{2\hbar})}
{\xi(u+2a-N-1+\frac{\pi i }{2\hbar})}.$
\end{lemma}
The following relation is a kind of 
 Hirota-Miwa equation\cite{H81},\cite{M82},  
which is a Pl\"{u}cker relation and used in a similar 
context \cite{KLWZ97,S00,DDT00,KOSY01}. 
\begin{lemma}\label{plucker}
$\xi^{(a)}_{m}(u)\xi^{(a)}_{m}(u+2)=
\xi^{(a)}_{m-1}(u)\xi^{(a)}_{m+1}(u+2)+
\xi^{(a-1)}_{m}(u)\xi^{(a+1)}_{m}(u+2)$.
\end{lemma}
From Lemma \ref{dual-Ca} and Lemma \ref{plucker}, 
we finally obtain: 
\begin{theorem}\label{main}
For $a \in I_{\sigma}$ and $m \in {\mathbb Z}_{\ge 1}$, 
$T^{(a)}_{m}(u)=\frac{\xi^{(a)}_{m}(u-a-m+1)}{\xi (u-a-m+1)}$
satisfies the  $T$-system for $U_{q}(A^{(2)}_{N})$ 
(\ref{t-sys1}), (\ref{t-sys2}).
\end{theorem}
There is another expression to the solution to the 
$U_{q}(A^{(2)}_{N})$ $T$-system (\ref{t-sys1}), (\ref{t-sys2}), 
which follows from a reduction of  
the Bazhanov and Reshetikhin's 
Jacobi-Trudi type formula \cite{BR90} (cf. section 5 in \cite{KS95-2}). 
\begin{eqnarray}
T^{(a)}_{m}(u)=\det_{1\le j,k \le m}({\mathcal T}^{a-j+k}(u+j+k-m-1)),
\label{jacobi-trudi}
\end{eqnarray}
where ${\mathcal T}^{a}(u)$ obeys the following condition:
\begin{eqnarray}
{\mathcal T}^{a}(u)=
\left\{
 \begin{array}{lll}
  0 & \mbox{if} & a <0 \; \mbox{or} \; a > N+1 \\
  1 & \mbox{if} & a=0 \; \mbox{or} \; a=N+1 \\
  T^{(a)}_{1}(u) & \mbox{if} & 1 \le a \le n \\ 
  T^{(N-a+1)}_{1}(u+\frac{\pi i}{2\hbar}) & \mbox{if} &
   n+1 \le a \le N .
 \end{array}
\right.
\end{eqnarray}
Through the identification ${\mathcal T}^{a}(u)=T^{a}(u)$ 
and Lemma \ref{m-nnsy}, 
(\ref{jacobi-trudi}) reproduces the solution in 
Theorem \ref{main}, and also 
the tableaux sum expression in \cite{KS95-2}. 
\section{Discussion}
In this paper, we have dealt with the $T$-system 
without the vacuum part. 
On applying our results to realistic problems 
in solvable lattice models or integrable field theories, 
we must specify the Baxter $Q$-function, and 
recover the vacuum part 
whose shape depends on each model. 
We can easily recover the vacuum part  
by multiplying the vacuum function $\psi_{a}(u)$ 
by the function $z_{a}(u)$ so that $\psi_{a}(u)$ is compatible with the 
Bethe ansatz equation of the form (cf. \cite{RW87,KOS95})
\begin{eqnarray}
\Psi_{a}(u^{(a)}_{j})=\prod_{b=1}^{n^{\prime}}
\frac{Q_{b}^{r_{ab}}(u^{(a)}_{j}+(\alpha_{a}|\alpha_{b}))}
{Q_{b}^{r_{ab}}(u^{(a)}_{j}-(\alpha_{a}|\alpha_{b}))},
\qquad a\in I_{\sigma}. \label{BAE}
\end{eqnarray}
In the case of the solvable vertex model,  it was 
 conjectured \cite{KOS95} that $\Psi_{a}(u)$ is given as a  
 ratio of  Drinfeld polynomials. 

A remarkable 
connection between DVF and the $q$-character was pointed out 
in \cite{FR99}. 
It was also conjectured \cite{KOSY01} that $q$-characters of 
 Kirillov-Reshetikhin modules over $U_{q}(X^{(1)}_{n})$ 
satisfy the $T$-system\cite{KNS95}. 
It is natural to expect that similar phenomena are
 also observed for the twisted case $U_{q}(X^{(r)}_{N})$ ($r>1$). 
 Thus one may look upon $T^{(a)}_{m}(u)$ in Theorem \ref{main} 
(or $T^{a}(u)$) 
as a kind of $q$-character. 
Precisely speaking, 
in view of a correspondence \cite{FR98} between DVF 
and generators of the deformed $W$-algebra, 
one may need to slightly modify $T^{(a)}_{m}(u)$  (or $T^{a}(u)$) 
(in particular, the factor $\frac{\pi i}{\hbar}$) 
to identify $T^{(a)}_{m}(u)$ (or $T^{a}(u)$) 
with the $q$-character of 
the Kirillov-Reshetikhin module over 
$U_{q}(X^{(r)}_{N})$ ($r>1$). 

We can also easily construct difference $L$ operators 
associated with superalgebras 
by using the results on the 
analytic Bethe ansatz \cite{T97,T98,T99-1,T99}. 
However their orders are infinite as 
$U_{q}(B^{(1)}_{n}),U_{q}(D^{(1)}_{n}),
U_{q}(D^{(2)}_{n+1}),U_{q}(D^{(3)}_{4})$ case. 
Thus we will need some new ideas 
to construct Casorati determinant like 
solutions to the $T$-system for superalgebras. 
\section*{Acknowledgments}
\noindent
The author would like to thank Professor 
A. Kuniba for explaining the results on \cite{KOSY01}. 
He is financially supported by 
Inoue Foundation for Science. 
  
\end{document}